\documentclass[12pt,preprint]{aastex}

\slugcomment{submitted to ApJ}

\shorttitle{A Substellar Companion to 11 Com}
\shortauthors{Liu et al.}

\begin{document}


\title{A Substellar Companion to the Intermediate-Mass Giant 11 Com}


\author{
Y.J. Liu,\altaffilmark{1,2} Bun'ei Sato,\altaffilmark{3} G.
Zhao,\altaffilmark{1} Kunio Noguchi,\altaffilmark{4} H.
Wang,\altaffilmark{1} Eiji Kambe,\altaffilmark{5} Hiroyasu
Ando,\altaffilmark{4} Hideyuki Izumiura,\altaffilmark{5,6} Y.Q.
Chen,\altaffilmark{1} Norio Okada,\altaffilmark{4} Eri
Toyota,\altaffilmark{7} Masashi Omiya,\altaffilmark{8} Seiji
Masuda,\altaffilmark{9} Yoichi Takeda,\altaffilmark{4} Daisuke
Murata,\altaffilmark{7} Yoichi Itoh,\altaffilmark{7} Michitoshi
Yoshida,\altaffilmark{5} Eiichiro Kokubo,\altaffilmark{4} and
Shigeru Ida\altaffilmark{3} }


 \altaffiltext{1}{National Astronomical Observatories, Chinese Academy of Sciences,
 A20 Datun Road, Chaoyang District, Beijing 100012, China;
lyj@bao.ac.cn, gzhao@bao.ac.cn, wh@bao.ac.cn, cyq@bao.ac.cn}
\altaffiltext{2}{Graduate University of Chinese Academy of Sciences,
Beijing 100049, China}
 \altaffiltext{3}{Tokyo Institute of Technology, 2-12-1 Ookayama, Meguro-ku,
Tokyo 152-8550, Japan; sato.b.aa@m.titech.ac.jp, ida@geo.titech.ac.jp}
 \altaffiltext{4}{National Astronomical Observatory of Japan,
National Institutes of Natural Sciences, 2-21-1 Osawa, Mitaka,
Tokyo 181-8588, Japan; knoguchi@optik.mtk.nao.ac.jp,
ando@optik.mtk.nao.ac.jp, okada@design.mtk.nao.ac.jp,
takedayi@cc.nao.ac.jp, kokubo@nao.ac.jp}
 \altaffiltext{5}{Okayama Astrophysical Observatory, National
Astronomical Observatory of Japan, National Institutes of Natural
Sciences, 3037-5 Honjyo, Kamogata, Asakuchi, Okayama 719-0232, Japan;
kambe@oao.nao.ac.jp, izumiura@oao.nao.ac.jp, yoshida@oao.nao.ac.jp}
 \altaffiltext{6}{The Graduate University for Advanced Studies,
Shonan Village, Hayama, Kanagawa 240-0193, Japan}
 \altaffiltext{7}{Graduate School of Science, Kobe University,
1-1 Rokkodai, Nada, Kobe 657-8501, Japan;
toyota@kobe-u.ac.jp, murata@harbor.scitec.kobe-u.ac.jp, yitoh@kobe-u.ac.jp}
 \altaffiltext{8}{Department of Physics, Tokai University,
1117 Kitakaname, Hiratsuka, Kanagawa 259-1292, Japan;
ohmiya@peacock.rh.u-tokai.ac.jp}
 \altaffiltext{9}{Tokushima Science Museum, Asutamu Land Tokushima,
45-22 Kibigadani, Nato, Itano-cho, Itano-gun, Tokushima 779-0111,
Japan; masuda@asutamu.jp}


\begin{abstract}
We report the detection of a substellar companion orbiting the
intermediate-mass giant star 11 Com (G8 III).
Precise Doppler measurements of the star
from Xinglong station and Okayama Astrophysical Observatory (OAO)
revealed Keplerian velocity variations with an orbital
period of 326.03$\pm$0.32 days, a semiamplitude of 302.8$\pm$2.6 m s$^{-1}$,
and an eccentricity of 0.231$\pm$0.005.
Adopting a stellar mass of 2.7$\pm$0.3 $M_{\odot}$, the minimum mass of the
companion is 19.4$\pm$1.5 $M_{\rm J}$, well above the deuterium burning limit,
and the semimajor axis is 1.29$\pm$0.05 AU.
This is the first result from the joint planet search program
between China and Japan aiming at revealing statistics of
substellar companions around intermediate-mass giants.
11 Com b emerged from 300 targets of the planet search
program at OAO. The current detection rate of a brown dwarf candidate
seems to be comparable to that around solar-type stars within
orbital separations of $\sim$3 AU.
\end{abstract}



\section{Introduction}
A ``brown dwarf desert'' is widely known as a deficit in the
frequency of brown dwarf companions ($13-80M_{\rm J}$) to
solar-type stars (e.g. Marcy \& Butler 2000). The mass distribution
of companions rises steeply toward planets at one extreme
and stellar masses at the other, with a deficit of brown dwarfs
in between (e.g. Grether \& Lineweaver 2006).
The presence of such a desert within orbital separation of
$\sim$3 AU was revealed and confirmed by the precise radial
velocity surveys during the past decade (Butler et al. 2006
and references therein), and recent coronagraphic surveys
showed that the desert extends to $\sim$1000 AU
(McCarthy \& Zuckerman 2004).
This bimodal mass distribution of the low-mass
companions is considered to be the outcome of distinct
formation mechanisms for planets and stellar companions
to solar-type stars.

On the other hand, it has remained to be cleared whether
the brown dwarf desert also exists around massive stars.
Few absorption lines in the spectra of massive early-type
dwarfs, which are often rotationally broadened,  make it
difficult to detect low-mass companions by precise radial
velocity measurements, and their high luminosity prevents
us from directly imaging close and faint companions.
Therefore, massive stars had not been major targets of
planet/brown dwarf searches compared with lower-mass stars.

However, the existence of substellar companions
to intermediate-mass stars ($\gtrsim1.6M_{\odot}$)
has been gradually uncovered mainly through the precise
Doppler surveys of G, K giants (e.g. Setiawan et al. 2005;
Sato et al. 2007; Hatzes et al. 2005; Johnson et al. 2007;
Lovis \& Mayor 2007; Frink et al. 2002; Niedzielski et al. 2007).
These are massive stars in evolved stages and have many sharp
absorption lines in their spectra appropriate for radial
velocity measurements.
The minimum masses of the discovered companions range
from 0.6 to 19.8 $M_{\rm J}$, and the semimajor axes are
from 0.8 to 2.4 AU. The two of the most massive ones,
NGC4349 No127 b (19.8 $M_{\rm J}$; Lovis \& Mayor 2007)
and HD 13189 b (14 $M_{\rm J}$; Hatzes et al. 2005)
\footnote{HD 13189 b has uncertainty between 8 to 20 $M_{\rm J}$
depending on the uncertainty of the host star's mass
(2--6 $M_{\odot}$) (Hatzes et al. 2005)},
are falling in the brown dwarf regime.
Galland et al. (2005) have tried
a precise Doppler survey of low-mass companions
to A--F dwarfs and found one brown dwarf candidate with
the minimum mass of 25$M_{\rm J}$ in a 25-day orbit
around the A9 V star HD 180777.
Kouwenhoven et al. (2007) found two distant (520 and 1500 AU)
brown dwarf candidates to a B star in the Sco OB2 association
by direct imaging with the near-infrared camera and the
adaptive optics system. They derived a frequency of brown
dwarf companions ($\gtrsim30M_{\rm J}$) to 0.5$\pm$0.5\%
in the separation range 130--520 AU for late-B and A type
stars in the association, which is comparable to the
frequency of 0.7$\pm$0.7\% derived by McCarthy \& Zuckerman
(2004) for F--M stars in the separation range 120--1200 AU.

We here report the detection of a substellar companion to the
intermediate-mass G-type giant 11 Com (HD 107383) from Doppler planet search
programs at the National Astronomical Observatories (Xinglong,
China) and Okayama Astrophysical Observatory (OAO, Japan). This is
the third one with the minimum mass falling in the brown dwarf regime
discovered around intermediate-mass giants.

\section{Observations}

\subsection{OAO Observations}
The Okayama Planet Search Program started in 2001 using
a 1.88 m telescope and the HIgh Dispersion Echelle Spectrograph
(HIDES; Izumiura 1999) at OAO. It aims to detect planets
around intermediate-mass G-type (and early K-type) giants
(Sato et al. 2005) and about 300 targets are now under survey.
For precise radial velocity measurements, we set a wavelength range
to 5000--6100${\rm \AA}$ and a slit width to 200 $\mu$m
($0.76^{\prime\prime}$) giving a spectral
resolution ($\lambda/\Delta\lambda$) of 67000,
and use an iodine absorption cell (I$_2$ cell; Kambe et al.
2002) for precise wavelength calibration.
Our modeling technique of an I$_2$-superposed stellar spectrum
(star+I$_2$) is detailed in Sato et al. (2002), which is based
on the method by Butler et al. (1996), giving
a Doppler precision of about 6 m s$^{-1}$ over a time span of 6 years.
The stellar template used for radial velocity analysis is extracted
from several star+I$_2$ spectra with use of the method described
in Sato et al. (2002).
The reduction of echelle data is performed using the IRAF
\footnote{IRAF is distributed by the National Optical Astronomy
Observatories, which is operated by the Association of Universities
for Research in Astronomy, Inc. under cooperative agreement with the
National Science Foundation, USA.} software package in the standard manner.
For abundance analysis of 11 Com, we took pure (I$_2$-free) stellar
spectra of wavebands of 5000--6100 and 6000--7100 ${\rm \AA}$
with the same wavelength resolution as that for radial velocity
measurements.

\subsection{Xinglong Observations}

The Xinglong Planet Search Program started in 2005 within a
framework of international collaboration between China and Japan
aiming to extend the planet search program ongoing at OAO. At the
National Astronomical Observatories (Xinglong, China), we have
mainly monitored about 100 G-type giants with a visual magnitude of
$V\sim6$, which have not been observed at OAO.

Before starting the planet search program, we installed an I$_2$
cell to the Coude Echelle Spectrograph (CES; Zhao \& Li 2001), which
attached to 2.16 m telescope at Xinglong Station in 2004 August. The
cell is a copy of those for HIDES at OAO and High Dispersion
Spectrograph at Subaru telescope (Kambe et al. 2002). The cell is
put in front of the entrance slit of the spectrograph and its
temperature is controlled to 60 $^\circ$C. We use a blue-arm and a
middle-camera configuration which covers a wavelength range of
3900--7260${\rm \AA}$ with a spectral resolution of $\sim40000$ by
two pixels sampling. Although a wide wavelength range can be
obtained with a single exposure, only a waveband of 470 ${\rm \AA}$
is available for radial velocity measurements due to the small
format (1K$\times$1K) of the current CCD, which is a thinned back illuminated
blue enhanced TEK CCD whose pixel size is 24$\times$24 $\mu$m$^{2}$.

The modeling technique of a star+I$_2$ spectrum and the extraction
method of a stellar template are based on Sato et al. (2002)'s code.
For CES data, we use 5 Gaussian
profiles to reconstruct the instrumental profile and the second
order Legendre polynomial function to express the wavelength
scale. Entire echelle spectrum is divided into about forty
segments (130--150 pixels for each), and Doppler analysis is
applied to each segment. The current best short-term ($\sim1$ week)
precision for bright stars is about 15 m s$^{-1}$ limited by the
low wavelength resolution and the narrow wavelength coverage of
the spectrograph. In longer time scale ($\sim$ 1 yr), the precision
gets worse to 20--25 m s$^{-1}$ including velocity offsets between
different observing runs.
In the case of fainter stars, the measurement precision is mainly
limited by a signal-to-noise ratio (S/N).
We can typically achieve a precision of 30--40 m s$^{-1}$ for a
$V=6$ star with a S/N$\sim$150 for an exposure time of 1800 sec.

\section{Stellar Properties}

11 Com (HR 4697, HD 107383, HIP 60202) is a G8 III giant star with a
$V$ magnitude $V=4.74$, a color index $B-V=0.99$, and a precise
astrometric parallax $\pi=9.04\pm0.86$ mas (ESA 1997). The resulting
distance is 112 pc from the sun and the absolute magnitude is
$M_V=-0.48$. Effective temperature $T_{\rm eff}=4742\pm100$K was
derived from the $B-V$ and metallicity [Fe/H] using the empirical
calibration of Alonso et al. (2001), and a bolometric correction
$B.C.=-0.36$ was derived from the calibration of Alonso et al.
(1999) depending on temperature and metallicity. Then the stellar
luminosity was estimated to $L=172\pm31L_{\odot}$, and a radius to
$R=19.5\pm2R_{\odot}$. Stellar mass $M=2.7\pm0.3M_{\odot}$ was
estimated from the star's position in the theoretical H-R diagram by
interpolating in the evolutionary tracks of Yonsei-Yale (Yi et al.
2003). Surface gravity $\log g=2.31\pm0.10$ was determined using the
relation between temperature, mass, luminosity and gravity (see
equation 1 of Chen et al. 2000). Iron abundance was determined from
the equivalent widths measured from the I$_2$-free spectrum
(5000--6100 and 6000--7100 ${\rm \AA}$) combined with the model
atmosphere (Kurucz 1993). We iterated the whole procedure described
above until the value of the overall metallicity of the model
converged. Finally we got ${\rm [Fe/H]}=-0.35\pm0.09$ and a
microturbulent velocity $v_t=1.5\pm0.2$ km s$^{-1}$. The rotational
velocity of the star is determined by de Medeiros \& Mayor (1999) to
be $v \sin i=1.2\pm1.0$ km s$^{-1}$. {\it Hipparcos} made a total of
99 observations of the star, revealing a photometric stability down
to $\sigma=0.006$ mag. The stellar parameters are summarized in
Table \ref{tbl-1}.

Figure 1 shows Ca {\small II} H line for 11 Com
together with those for G8--9 III giants in our sample.
Although the calibration to measure chromospheric activity
for our spectra and the correlation between chromospheric activity
and intrinsic radial velocity ``jitter'' for giants have not
been well established yet,
the lack of significant emission in the line of 11 Com
suggests its low chromospheric activity.

\section{Radial Velocities and Orbital Solution}
We started observations of 11 Com at OAO in 2003 December.
Soon after that, its large radial velocity variations were detected
suggesting the existence of a substellar companion.
Based on this result, we started observations of this star
at Xinglong in 2005 to check the measurement precision with
CES and to confirm the variations of the star independently.
Until February 2007, we gathered a total of 28 data points of the
star at OAO with a typical S/N of 250 per pixel, and a total
of 18 data points at Xinglong with a typical S/N of 150--200.

The observed radial velocities are shown in Figure \ref{fig2} and
listed in Table \ref{tbl-2} (OAO) and \ref{tbl-3} (Xinglong)
together with their estimated uncertainties. The best-fit Keplerian
orbit was derived using both of the OAO and Xinglong data.
An offset of $-$205 m s$^{-1}$ was applied to the Xinglong radial
velocities in order to minimize $\chi^2$ when fitting a Keplerian
model to the combined OAO and Xinglong velocities.
The resulting orbit is shown in Figure \ref{fig2}
overplotted on the velocities, and its parameters are listed in
Table \ref{tbl-4}. The uncertainty of each parameter was estimated
using a Monte Carlo approach. The radial velocity variability can be
well fitted by a Keplerian orbit with a period $P=326.03\pm0.32$
days, a velocity semiamplitude $K_1=302.8\pm2.6$ m s$^{-1}$, and an
eccentricity $e=0.231\pm0.005$. When we determined a Keplerian orbit
using only OAO data, the rms scatter of the residuals to the Keplerian fit
for OAO data was 17.8 m s$^{-1}$, which is slightly larger
than the typical intrinsic radial velocity scatter of late G-type
giants in our sample (Sato et al. 2005), but we found
no significant additional periodicity in the residuals at this
stage. The Keplerian orbit derived by only OAO data fits the Xinglong
data with the rms scatter of 33.2 m s$^{-1}$, which is comparable to
the observational errors.
Adopting a stellar mass of 2.7$\pm$0.3 $M_{\odot}$, we obtained for the
companion a mass $m_2\sin i=19.4\pm1.5$ $M_{\rm J}$ and a semimajor
axis $a=1.29\pm0.05$ AU. The uncertainties mostly come from that in
host star's mass. If we assume the orbit is randomly oriented, there
is a 3\% chance that the true mass exceeds 80 $M_{\rm J}$
($i<14^{\circ}$), a border between brown dwarf and star mass regimes.

To investigate other causes producing apparent radial velocity
variations such as pulsation and rotational modulation,
spectral line shape analysis was performed with use of high
resolution stellar templates followed by the technique of
Sato et al. (2007). We extracted two stellar templates from the
star+I$_2$ spectra obtained at OAO: one was from five spectra
with observed radial velocities of $250\sim300$ m s$^{-1}$
and the other was from those with $\sim-200$ m s$^{-1}$.
Cross correlation profiles of the templates were calculated
for 50 spectral segments (4--5${\rm \AA}$ width each) in which
severely blended lines or broad lines were not included.
Three bisector quantities were calculated for the cross correlation
profile of each segment: the velocity span (BVS), which is the velocity
difference between two flux levels of the bisector; the velocity
curvature (BVC), which is the difference of the velocity span of the
upper half and lower half of the bisector; and the velocity displacement
(BVD), which is the average of the bisector at three different flux levels.
We used flux levels of 25\%, 50\%, and 75\% of the cross correlation
profile to calculate the above quantities. Figure \ref{fig3}
shows the resulting
bisector quantities plotted against the center wavelength of each segment.
As expected from the planetary hypothesis, both of the bisector velocity
span and the curvature are identical to zero ($-$2.3 m s$^{-1}$ and
$-$2.4 m s$^{-1}$ in average, respectively), which means that the cross
correlation profiles are symmetric, and the average bisector velocity
displacement of $-$453.0 m s$^{-1}$ is consistent with the velocity
difference between the two templates.
Furthermore, these values are irrelevant to wavelength.
Based on these results, we conclude that the radial velocity variability
observed in 11 Com is best explained by orbital motion.

\section{Discussion and Summary}
We reported a discovery of a brown dwarf mass companion
to the intermediate-mass giant 11 Com. This is the first result
from the joint planet search program between China and Japan.
We obtained radial velocities of the star over a time span
of 3 years at OAO and 2 years at Xinglong. The two sites
independently detected the velocity variability of the star,
and the agreement of the two data sets assures the reliability
of our discovery and the radial velocity measurements at both sites.

11 Com b is the third substellar companion
ever discovered to intermediate-mass giants falling in the
brown dwarf regime (Fig \ref{fig4}).
To date, 10 planetary mass companions ($0.6-10M_{\rm J}$)
have been found around intermediate-mass ($\gtrsim1.6M_{\odot}$)
giants by precise Doppler surveys, but only 3 brown dwarf mass ones
($\gtrsim14M_{\rm J}$) have been found despite the larger radial
velocity amplitudes being easily detectable.
These results may suggest a shortage of brown dwarfs
relative to planets around intermediate-mass giants
within orbital separations of $\sim$3 AU.

Brown dwarf companions are generally thought to form by
gravitational collapse in protostellar clouds like
stellar binary systems (Bonnell \& Bastien 1992; Bate 2000).
It is difficult, however, to form close binary
systems with large difference in mass between the primary
and the secondary (Bate 2000). According to this scenario,
fewer brown dwarf companions are expected in higher-mass
systems than in lower-mass ones.
Around solar-type stars, the frequency of brown dwarf
companions is estimated to be less than 0.5\% within
3 AU (Marcy \& Butler 2000)
\footnote{Patel et al. (2007) recently discovered five brown
dwarf candidates beyond 3 AU by precise radial velocity technique.}.
As long as our survey, 11 Com b emerged from a sample of
300 stars at OAO. The current detection rate seems to be comparable
to that around solar-type stars, although the exact detection limits
for all the targets in our sample have not been derived yet.
Detailed statistical analysis will be presented in a forthcoming paper.
Ongoing large scale Doppler surveys of G--K giants (Setiawan et al.
2005; Sato et al. 2007; Hatzes et al. 2005; Johnson et al. 2007;
Lovis \& Mayor 2007; Frink et al. 2002; Niedzielski et al. 2007)
and B--A dwarfs (Galland et al. 2005) around the globe
will derive statistically reliable value of the frequency.

Gravitational instability in protostellar disks is
another scenario to form brown dwarf companions (Boss 2000;
Rice et al. 2003).
This scenario is also proposed as the formation mechanism
for massive gaseous planets.
A half of the planetary mass companions discovered around
intermediate-mass stars have minimum masses greater
than 5 $M_{\rm J}$,
so-called ``super planets'' (Hatzes et al. 2005;
Sato et al. 2003, 2007; Setiawan et al. 2003, 2005;
Lovis \& Mayor 2007).
Can these companions and 11 Com b be put into
the same context in the point of formation mechanism?
One remarkable feature of these companions is that most of
them (including 11 Com b) are orbiting stars with subsolar
metallicity. This apparently favors gravitational instability
scenario as their formation mechanism (Boss 2002), which is
less sensitive to metallicity than core accretion scenario is
(Fischer \& Valenti 2005; Ida \& Lin 2004).
However, if the surface density of dust in the disk scales
as $M_{*}^{\alpha}Z$ ($\alpha> 0$), where $M_{*}$ is the stellar
mass and $Z$ is the metallicity, massive stars may have higher
disk surface densities, which may compensate for lower metalicities.
Such disks could form massive solid cores that can accrete huge gas
envelope up to 20$M_{\rm J}$.
Further investigation of mass and metallicity distribution of
substellar companions to massive stars helps discriminate between
these formation scenarios.

\acknowledgments

This research is based on data collected at Xinglong station, which
is operated by the National Astronomical Observatories, CAS (NAOC),
and Okayama Astrophysical Observatory (OAO), which is operated by
National Astronomical Observatory of Japan (NAOJ). We thank Junjun
Jia, Jiaming Ai, and Hongbin Li for their expertise and support of
the Xinglong observations. We are grateful to all of the staffs of
OAO for their support during the OAO observations. Data analysis was
in part carried out on ``sb'' computer system operated by the
Astronomical Data Analysis Center (ADAC) and Subaru Telescope of
NAOJ. We thank the National Institute of Information and
Communications Technology for their support on high-speed network
connection for data transfer and analysis. This work was funded by
the National Natural Science Foundation of China under grants
10521001 and 10433010. B.S., H.I., H.A., and M.Y. are supported by
Grants-in-Aid for Scientific Research No.17740106, (C) No.13640247,
(B) No.17340056, (B) No.18340055, respectively, from the Japan
Society for the Promotion of Science (JSPS). E.T., D.M., and Y.I. are
supported by ``The 21st Century COE Program: The Origin and
Evolution of Planetary Systems'' in Ministry of Education, Culture,
Sports, Science and Technology (MEXT). E.K. and S.I. are partially
supported by MEXT, Japan, the Grant-in-Aid for Scientific Research
on Priority Areas, ``Development of Extra-Solar Planeary Science.''
This research has made use of
the SIMBAD database, operated at CDS, Strasbourg, France.

\clearpage



\begin{figure}
\epsscale{.80}
\plotone{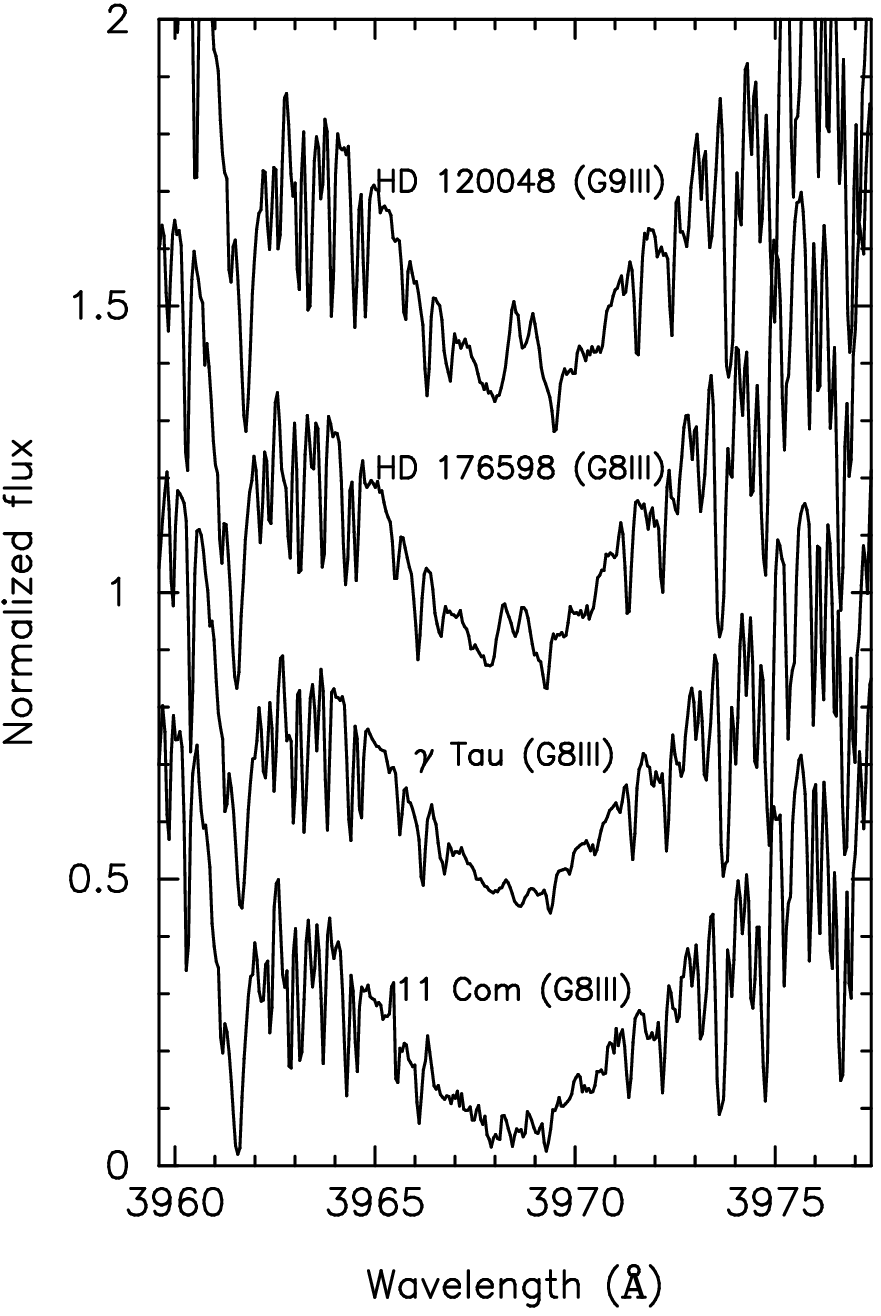}
\caption{Spectra in the region of CaII H lines.
HD 120048 and HD 176598 show significant core reversals
in the lines suggesting high chromospheric activity.
They have relatively large radial velocity scatters of 25--35
m s$^{-1}$. $\gamma$ Tau exhibits no emission in the line and show
radial velocity stability at a level of $\sim$5 m s$^{-1}$.\label{fig1}}
\end{figure}

\begin{figure}
\epsscale{.80}
\plotone{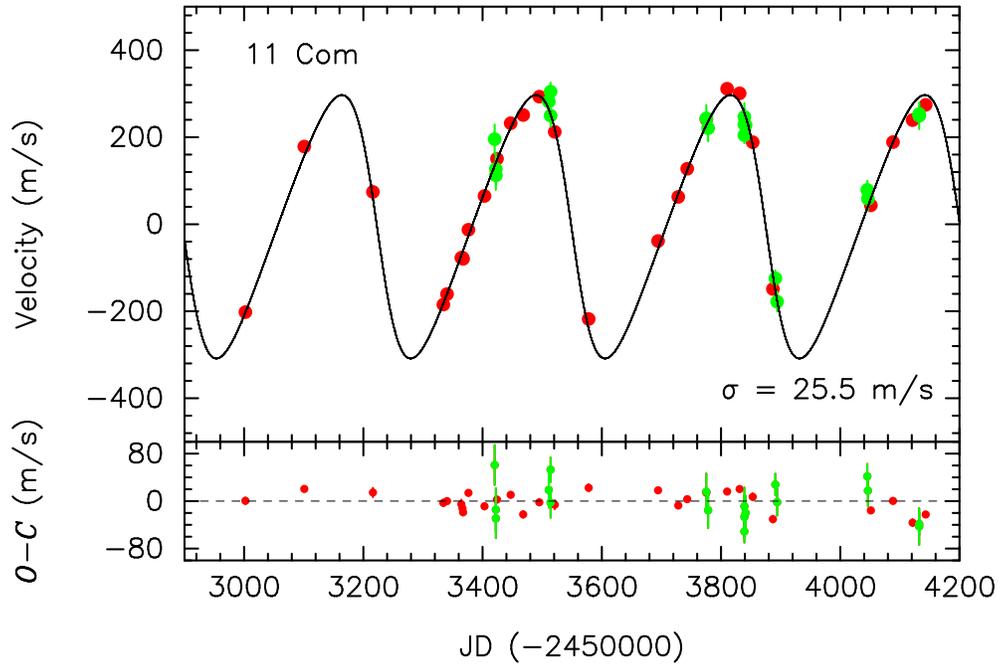}
\caption{Radial velocities of 11 Com (dots) observed
at OAO (red) and Xinglong (green).
The Keplerian orbit (solid line) is determined using
both of the OAO and Xinglong data.
\label{fig2}}
\end{figure}

\begin{figure}
\plotone{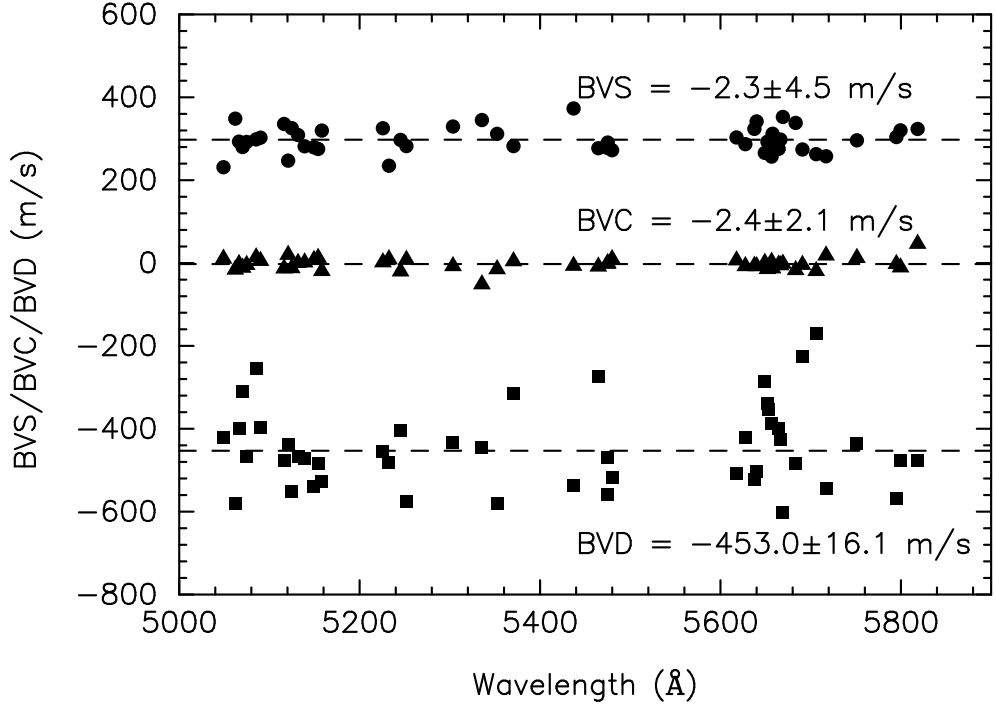}
\caption{Bisector quantities of cross-correlation profiles
between the templates of 11 Com at peak (250--300 m s$^{-1}$)
and valley ($-$200 m s$^{-1}$) phases of observed radial
velocities, showing bisector velocity span (BVS, circle),
bisector velocity curvature (BVC, triangle),
and bisector velocity displacement (BVD, square). The
definition of these quantities are described in Section 4.
The mean values and their standard errors are shown in the
figure. Dashed lines represent mean values of the quantities
(an offset of 300 m s$^{-1}$ is added to the BVS).
\label{fig3}}
\end{figure}

\begin{figure}
\plotone{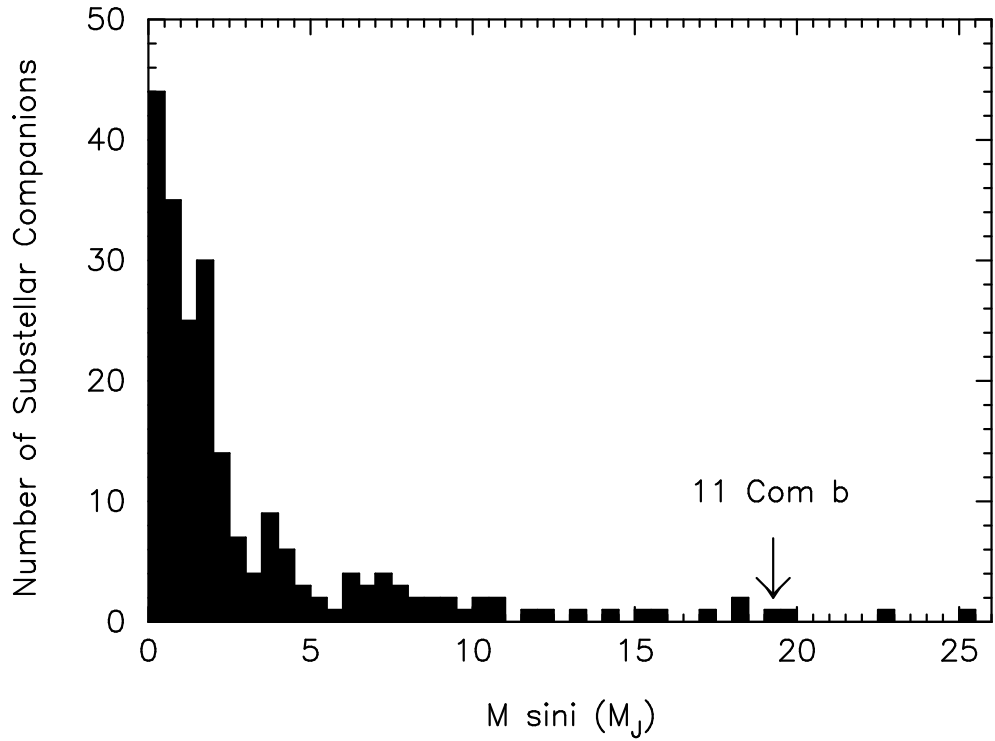}
\caption{Mass distribution of substellar companions
discovered by precise Doppler surveys
(including all of the host stars with $\lesssim4M_{\odot}$).
The data are based on the list from
http://exoplanets.org/planets.shtml.
NGC2423 No3 b (10.6 $M_{\rm J}$), NGC4349 No127 b (19.8 $M_{\rm J}$),
HD 13189 b (K2 II, 14$M_{\rm J}$), HD 17092 b (K0 III, 4.6 $M_{\rm J}$) and
HD 180777 b (A9 V, 25$M_{\rm J}$)
are not included in the list but are added to the histogram.
\label{fig4}}
\end{figure}







\clearpage

\begin{deluxetable}{lr}
\tabletypesize{\small}
\tablecaption{Stellar Parameters for 11 Com\label{tbl-1}}
\tablewidth{0pt}
\tablehead{
\colhead{Parameter} & \colhead{Value}}
\startdata
Sp. Type       & G8 III \\
$\pi$ (mas)    & 9.04$\pm$0.86 \\
$V$            & 4.74   \\
$B-V$          & 0.99   \\
$M_{V}$        & $-$0.48 \\
$B.C.$         & $-$0.36 \\
$T_{\rm eff}$ (K) & 4742$\pm$100 \\
$\log g$       &  2.31$\pm$0.1 \\
$[$Fe/H$]$     &  $-$0.35$\pm$0.09 \\
$v_{\rm t}$ (km s$^{-1}$)   & 1.5$\pm$0.2\\
$L$ ($L_{\odot}$) &  175$\pm$31  \\
$R$ ($R_{\odot}$) &   19$\pm$2  \\
$M$ ($M_{\odot}$) &  2.7$\pm$0.3  \\
$v\sin i$ (km s$^{-1}$) & 1.2$\pm$1.0\\
\enddata

\end{deluxetable}

\begin{deluxetable}{ccc}
\tabletypesize{\small}
\tablecaption{Observed Radial Velocities of 11 Com at OAO\label{tbl-2}}
\tablewidth{0pt}
\tablehead{
\colhead{JD} & \colhead{Radial Velocity}   & \colhead{Error}\\
\colhead{($-$2,450,000)} & \colhead{(m s$^{-1}$)}   & \colhead{(m s$^{-1}$)}
}
\startdata
3002.2791 & $-$201.7 & 5.7\\
3101.0241 & 178.7 & 4.6\\
3216.0035 & 74.5 & 7.9\\
3334.3172 & $-$184.5 & 5.5\\
3340.3401 & $-$160.2 & 4.6\\
3364.3191 & $-$76.9 & 7.9\\
3366.2249 & $-$76.6 & 5.2\\
3367.2333 & $-$79.3 & 5.4\\
3376.3243 & $-$12.5 & 4.7\\
3403.2636 & 65.3 & 5.1\\
3424.1132 & 150.7 & 5.1\\
3447.1803 & 232.5 & 4.6\\
3468.0504 & 251.2 & 5.6\\
3495.1120 & 293.4 & 4.5\\
3521.0884 & 212.5 & 7.5\\
3577.9651 & $-$217.7 & 6.2\\
3694.3575 & $-$38.2 & 5.0\\
3728.3204 & 63.2 & 5.1\\
3743.3484 & 127.7 & 5.0\\
3775.2089 & 241.6 & 4.7\\
3810.1629 & 311.6 & 3.9\\
3831.1262 & 301.0 & 4.5\\
3853.1393 & 188.7 & 5.8\\
3887.0520 & $-$148.5 & 5.7\\
4051.3576 & 43.8 & 4.6\\
4088.3683 & 189.0 & 4.8\\
4121.2706 & 239.6 & 5.9\\
4143.2374 & 274.5 & 4.0\\
\enddata
\end{deluxetable}

\begin{deluxetable}{ccc}
\tabletypesize{\small}
\tablecaption{Observed Radial Velocities of 11 Com at Xinglong\label{tbl-3}}
\tablewidth{0pt}
\tablehead{
\colhead{JD} & \colhead{Radial Velocity}   & \colhead{Error}\\
\colhead{($-$2,450,000)} & \colhead{(m s$^{-1}$)}   & \colhead{(m s$^{-1}$)}
}
\startdata
3420.2939 & 195.4 & 33.5\\
3422.2517 & 126.7 & 36.2\\
3422.2627 & 112.0 & 33.1\\
3511.0423 & 281.9 & 29.9\\
3514.0647 & 249.4 & 25.6\\
3514.0779 & 304.8 & 20.6\\
3775.2678 & 243.1 & 30.6\\
3778.3433 & 220.6 & 30.3\\
3839.1013 & 229.1 & 36.6\\
3839.1229 & 204.4 & 18.6\\
3839.1462 & 246.7 & 31.8\\
3841.1569 & 227.5 & 28.9\\
3891.0521 & $-$124.4 & 18.3\\
3894.0998 & $-$178.1 & 22.7\\
4045.3959 & 78.7 & 21.1\\
4046.3706 & 58.3 & 26.7\\
4132.3744 & 249.1 & 30.9\\
4132.3976 & 253.2 & 19.7\\
\enddata

\end{deluxetable}

\begin{deluxetable}{lr}
\tabletypesize{\small}
\tablecaption{Orbital Parameters of 11 Com determined from
both of OAO and Xinglong data.
\label{tbl-4}}
\tablewidth{0pt}
\tablehead{
\colhead{Parameter} & \colhead{Value}}
\startdata
$P$ (days)                    & 326.03$\pm$0.32\\
$K_1$ (m s$^{-1}$)            & 302.8$\pm$2.6\\
$e$                           & 0.231$\pm$0.005\\
$\omega$ (deg)                & 94.8$\pm$1.5\\
$T_p$    (JD$-$2,450,000)     & 2899.6$\pm$1.6\\
$a_1\sin i$ (10$^{-3}$AU)     & 8.848$\pm$0.073\\
$f_1(m)$ (10$^{-7}M_{\odot}$) & 8.68$\pm$0.21\\
$m_2\sin i$ ($M_{\rm J}$)     & 19.4$\pm$1.5\\
$a$ (AU)                      & 1.29$\pm$0.05\\
$N$                           & 46\\
rms (m s$^{-1}$)              & 25.5\\
Reduced $\sqrt{\chi^2}$       & 2.8\\
\enddata

\end{deluxetable}

\end{document}